\documentclass[twocolumn,aps,prd,10pt,superscriptaddress,longbibliography,floatfix,citeautoscript]{revtex4-2}

\usepackage{amsmath}
\usepackage{amssymb}
\usepackage{bm}
\usepackage{dcolumn}
\usepackage{glossaries}
\usepackage{graphicx}
\usepackage{multirow}
\usepackage{fancyvrb}
\usepackage[linesnumbered,ruled]{algorithm2e}
\usepackage{siunitx}
\usepackage[usenames,dvipsnames,svgnames]{xcolor}
\usepackage{hyperref}
\hypersetup{
    pdfnewwindow=true,      
    colorlinks=true,        
    linkcolor=Blue,         
    citecolor=Blue,         
    filecolor=Blue,         
    urlcolor=Blue           
}

\newacronym{dft}{DFT}{density functional theory}
\newacronym{md}{MD}{molecular dynamics}
\newacronym{mlp}{MLP}{machine-learned potential}
\newacronym{nep}{NEP}{neuroevolution potential}
\newacronym{rmse}{RMSE}{root mean square error}
\newacronym{reaxff}{ReaxFF}{reactive force-field} 
\newacronym{gpumd}{GPUMD}{Graphics Processing Units Molecular Dynamics}
\newacronym{mae}{MAE}{mean absolute error}
\newacronym{ann}{ANN}{artificial neural network}
\newacronym{snes}{SNES}{separable natural evolution strategy}
\newacronym{go}{GO}{graphene oxide}
\newacronym{rgo}{rGO}{reduced graphene oxide}
\newacronym{hnemd}{HNEMD}{homogeneous nonequilibrium molecular dynamics}
\newacronym{ptm}{PTM}{polyhedral template matching}
\newacronym{rmsd}{RMSD}{root mean square deviation}
\newacronym{xps}{XPS}{X-ray photoelectron spectroscopy}
\newacronym{qhp}{qHP}{quasi-hexagonal phase}
\newacronym{qtp}{qTP}{quasi-tetragonal phase}
\newacronym{2d}{2D}{two-dimensional}

\DeclareSIUnit\angstrom{\text{Å}}
\DeclareSIUnit{\atom}{atom}
\DeclareSIUnit{\step}{step}
\DeclareSIUnit{\atomstepsecond}{\atom\step\per\second}

\begin{document}

\title{Thermal conductivities of monolayer graphene oxide from machine learning molecular dynamics simulations}

\author{Bohan Zhang}
\thanks{These authors contributed equally to this work.}
\affiliation{College of Physical Science and Technology, Bohai University, Jinzhou 121013, P. R. China}

\author{Biyuan Liu}
\thanks{These authors contributed equally to this work.}
\affiliation{Department of Mechanical and Aerospace Engineering, Hong Kong University of Science and Technology, Clear Water Bay, Kowloon, Hong Kong}

\author{Penghua Ying}
\email{hityingph@tauex.tau.ac.il}
\affiliation{Department of Physical Chemistry, School of Chemistry, Tel Aviv University, Tel Aviv, 6997801, Israel}
\affiliation{Laboratory for multiscale mechanics and medical science, SV LAB, School of
Aerospace, Xi'an Jiaotong University, Xi'an 710049, China}

\author{Zherui Chen}
\affiliation{Future Technology School, Shenzhen Technology University, Shenzhen 518118, P. R. China}
\affiliation{College of Applied Sciences, Shenzhen University, Shenzhen 518060, P. R. China}

\author{Yanzhou Wang}
\affiliation{School of Electronic Engineering, Chengdu Technological University, Chengdu 611730, China}
\affiliation{Department of Applied Physics, Aalto University, FIN-00076 Aalto, Espoo, Finland}

\author{Yonglin Zhang}
\affiliation{Department of Mechanical and Aerospace Engineering, Hong Kong University of Science and Technology, Clear Water Bay, Kowloon, Hong Kong}

\author{Haikuan Dong}
\affiliation{College of Physical Science and Technology, Bohai University, Jinzhou 121013, P. R. China}

\author{Jinglei Yang}
\email{maeyang@ust.hk}
\affiliation{Department of Mechanical and Aerospace Engineering, Hong Kong University of Science and Technology, Clear Water Bay, Kowloon, Hong Kong}
\affiliation{HKUST Shenzhen-Hong Kong Collaborative Innovation Research Institute, Futian, Shenzhen, China}

\author{Zheyong Fan}
\email{brucenju@gmail.com}
\affiliation{College of Physical Science and Technology, Bohai University, Jinzhou 121013, P. R. China}

\date{\today}

\begin{abstract}
Graphene oxide (GO) exhibits rich chemical heterogeneity that strongly influences its structural, thermal, and mechanical properties, yet quantitatively linking reduction chemistry to heat transport remains challenging. In this work, we develop a machine-learned neuroevolution potential (NEP) trained on an existing density functional theory dataset (\textit{Angew.\ Chem.\ Int.\ Ed.}, \textbf{63} , e202410088 (2024)), achieving reasonable accuracy at a computational cost much lower than the existing machine-learned and empirical potentials. Leveraging this potential, we perform large-scale molecular dynamics (MD) simulations to model the thermal reduction of GO across realistic structural domains. Using the homogeneous nonequilibrium MD method with a proper quantum-statistical correction scheme, we find that reduced GO exhibits strongly suppressed thermal conductivities, ranging from a few to tens of Wm$^{-1}$K$^{-1}$, substantially lower than pristine GO without defects and far below graphene. Moreover, the thermal conductivity of reduced GO increases moderately with increasing OH/O ratio, except at the highest oxidation level (O/C=0.5) where this trend inverts, while decreasing significantly with increasing O/C ratio, a trend strongly correlated with the fraction of recovered graphene-like structures. Our work provides a computationally tractable and predictive atomistic machine learning framework for exploring how chemical structure governs heat transport in heterogeneous carbon materials.
\end{abstract}

\maketitle

\section{Introduction}
The physicochemical properties of \gls{go} are determined by the interactions between graphene and various oxygen functional groups, including epoxides, hydroxyls, and carbonyls~\cite{dreyer2010chemistry}. These distinct chemical features give rise to unique mechanical, thermal, and transport characteristics, rendering \gls{go} a critical material for thermal management, membrane separation, and energy storage~\cite{chua2014chemical}. Despite its practical importance, a quantitative understanding of how chemical reduction influences heat transport in \gls{go} remains inadequate. Structural complexities, such as diverse functional groups, local bonding configurations, and defect topologies, introduce strong phonon-scattering pathways, making the heat transfer mechanisms difficult to accurately quantify~\cite{khine2022functional}. Consequently, establishing the fundamental relationship between reduction chemistry and thermal conductivity in \gls{go} remains a prominent and unresolved challenge in the field.

Over the past decade, experimental research has demonstrated that the thermal conductivities of \gls{go} and \gls{rgo} depend heavily on the degree of oxidation and reduction~\cite{renteria2015strongly}. Early experiments on \gls{go} films reported in-plane thermal conductivities of only a few Wm$^{-1}$K$^{-1}$, confirming the strong phonon-scattering effect of oxygen functional groups~\cite{jin2015effects}. Due to complex microstructural reconstruction during reduction~\cite{renteria2015strongly}, experimentally pinpointing the precise influence of the oxygen-to-carbon (O/C) and the hydroxyl-to-epoxide (OH/O) ratios remains difficult. The stochastic nature of reduction introduces significant structural uncertainty, making it impossible to experimentally control these ratios independently. Consequently, theoretical predictions are required to disentangle these chemical variables and accurately quantify their individual impacts on thermal transport.

Classical \gls{md} provides a powerful framework for probing phonon scattering, defect interactions, and heat transport in structurally disordered materials such as \gls{go} at experimentally relevant length scales. However, the predictive accuracy of \gls{md} is fundamentally limited by the fidelity of the underlying interatomic potential. Widely used empirical force fields, such as \gls{reaxff}~\cite{van2001reaxff}, often fail to capture the diverse and evolving carbon-oxygen bonding environments inherent to \gls{go} reduction. This limitation may lead to inconsistent thermal conductivity predictions and inaccurate descriptions of oxygen functional-group energetics ~\cite{zou2016phonon, diao2017reactive}. Consequently, these deficiencies introduce substantial uncertainty in modeling chemistry-transport relationships, underscoring the critical need for interatomic potentials that balance high accuracy with the computational efficiency required for reactive C/H/O systems. 

\Glspl{mlp} have recently emerged as a powerful alternative to empirical force fields to drive \gls{md} simulations, offering near-\gls{dft} accuracy with far lower cost~\cite{behler2007prl, bartok2010gaussian}. Notably, the MACE architecture~\cite{batatia2022mace} has been successfully applied to investigate the thermal reduction mechanisms~\cite{El-Machachi24accelarated} and mechanical properties~\cite{El-Machachi24accelarated} of \gls{go}. However, despite these successes, the MACE architecture remains computationally intensive, making it unsuitable for thermal conductivity calculations that require extensive molecular dynamics sampling~\cite{dong2024jap}. To accurately model the O/C- and OH/O-ratio-dependent thermal transport in \gls{rgo}, which demands long simulation trajectories, a reactive yet highly scalable interatomic potential is essential.

Here, we introduce NEP-GO, a specialized \gls{nep} model developed for \gls{go}. Our model is built upon the state-of-the-art \gls{nep} framework~\cite{fan2021neuroevolution, fan2022gpumd, song2024general}, which has been widely adopted in modeling complex  materials~\cite{dong2024jap,ying2025advances}. By leveraging this efficient architecture, NEP-GO achieves computational speeds and system sizes several orders of magnitude faster than MACE, given a consistent reference dataset. This performance breakthrough allows for the construction of a transferable and efficient model for \gls{go}, facilitating the large-scale atomistic simulations essential for capturing thermal reduction processes and their impact on thermal conductivity.

\section{Methods}
\subsection{NEP training}
We constructed a \gls{mlp} for \gls{go} and \gls{rgo} employing the \gls{nep} framework implemented in the \textsc{gpumd} package (version 4.3)~\cite{xu2025gpumd}. The \gls{nep} approach, specifically designed for efficient \gls{md} simulations, adopts the atomic environment descriptor methodology established by Behler and Parrinello~\cite{behler2007prl}. In this scheme, the site energy of an atom is modeled via an \gls{ann} using a high-dimensional descriptor vector invariant to translation, rotation, and permutation of atoms of the same species. The trainable parameters in the model are optimized using the \gls{snes}~\cite{schaul2011high}.

Training data were sourced from the \gls{dft} dataset generated by El-Machachi \textit{et al.}~\cite{El-Machachi24accelarated}, comprising 3,816 structures. We only excluded three isolated-atom configurations that are not needed for the \gls{nep} method. The \gls{nep} model architecture was established with a radial cutoff of \SI{4.2}{\angstrom} and an angular cutoff of \SI{3.7}{\angstrom}. These relatively short cutoffs were partially motivated by the choice in the previous MACE model~\cite{El-Machachi24accelarated}, which used a cutoff of \SI{3.7}{\angstrom}. The descriptors employ nine radial functions and seven angular radial functions, each expressed as a linear combination of nine basis functions. The angular components include three-body descriptors up to $l=4$ and four-body descriptors up to $l=2$ in the spherical harmonics expansion. The fitting neural network comprises a single hidden layer with 50 neurons. The model was trained for 200,000 steps using a full-batch training strategy. The loss function is a weighted sum of \glspl{rmse} of total energy, atomic forces, and virial tensors, as well as regularization terms. 

\subsection{Thermal reduction simulations}
We employed the trained \gls{nep} model to perform large-scale \gls{md} simulations of the thermal reduction of \gls{go} to \gls{rgo}. This process involves the transformation and eventual removal of oxygen-containing functional groups, accompanied by the release of gaseous byproducts such as H$_2$O, CO$_2$, and CO. 

To generate the initial structures, we sampled the $N$-dimensional parameter space $\mathbf{P}=[p_1, \dots, p_N]$ defined in Ref.~\citenum{El-Machachi24accelarated}. The parameters determining the composition and arrangement of functional groups were selected as: (1) the oxygen-to-carbon ratio (O/C); (2) the hydroxyl fraction (OH/O), defined as the proportion of oxygen atoms belonging to hydroxyl groups; and (3) the ratio of functionalized edges to hydrogen-terminated edges.

For the simulations in this work, the fraction of functionalized edges was consistently set to zero (purely hydrogen-terminated edges). We generated two distinct series of \gls{go} structures. In the first series, the O/C ratio was fixed at 0.4 while the hydroxyl fraction (OH/O) was systematically varied from 0.1 to 0.5, corresponding to parameter vectors $\mathbf{P}=[0.4, 0.1-0.5, 0]$. In the second series, the OH/O ratio was held constant at 0.3 while the O/C ratio was varied from 0.1 to 0.5, corresponding to vectors $\mathbf{P}=[0.1-0.5, 0.3, 0]$. The resulting models contain 11033 to 17395 atoms. These structures capture key characteristics of \gls{go}, including topological disorder in the carbon backbone (e.g., non-hexagonal rings), while ensuring all carbon atoms remain threefold-coordinated without large pores or vacancies.

All \gls{md} simulations were performed using the \textsc{gpumd} package (version 4.3)~\cite{xu2025gpumd} with an integration time step of \SI{0.1}{\femto\second}. \gls{rgo} structures were generated through a sequential equilibration, heating, annealing, and quenching protocol, with temperature control using the Langevin thermostat~\cite{Bussi2007accurate} with a damping parameter of \SI{50}{\femto\second}. 

To determine the optimal reduction conditions, the thermal reduction process was initially evaluated at temperatures of \SI{700}{\kelvin}, \SI{800}{\kelvin}, \SI{900}{\kelvin}, and \SI{1200}{\kelvin}. We observed that annealing at \SI{900}{\kelvin} yields structurally stable \gls{rgo}, achieving an appropriate degree of reduction while preserving sufficient mechanical integrity. Consequently, \SI{900}{\kelvin} was selected as the annealing temperature for all subsequent simulations. This is quantitatively different from  Ref.~\cite{El-Machachi24accelarated}, which employed a reduction temperature of \SI{1500}{\kelvin}; we adopted \SI{900}{\kelvin} specifically to avoid over-reduction and maintain a moderate residual oxygen content.

The \gls{rgo} generation protocol proceeded as follows: First, each \gls{go} configuration was equilibrated at \SI{300}{\kelvin} for \SI{10}{\pico\second}, and then rapidly heated to the target reduction temperature of \SI{900}{\kelvin} over \SI{0.1}{\nano\second}, corresponding to a heating rate of \SI{6}{\kelvin\per\pico\second}. Following this, the reduction and annealing stages were carried out at constant temperature for \SI{1.9}{\nano\second}. Finally, the system was quenched to \SI{300}{\kelvin} within \SI{0.1}{\nano\second} and further equilibrated at this temperature for \SI{0.1}{\nano\second}. Throughout the entire process, detached fragments and gaseous byproducts were periodically removed to preserve the integrity of the graphene sheet. To account for the stochastic nature of the thermal reduction process, five independent simulations were performed for each specific parameter $\mathbf{P}$.

\subsection{Thermal conductivity calculations}
Thermal transport simulations were performed using the \gls{hnemd} method~\cite{Fan2019prb} within the canonical (NVT) ensemble employing the Bussi-Donadio-Parrinello thermostat~\cite{bussi2007canonical}, as implemented in the \textsc{GPUMD} package (version 4.3)~\cite{xu2025gpumd}. For each case, a production run of \SI{3}{\nano\second} was performed following \SI{0.2}{\nano\second} equilibration at \SI{300}{\kelvin}. 

In the \gls{hnemd} method, a steady-state heat current is induced by applying an external driving force $\mathbf{F}_{i}^{\rm ext}$ to each atom $i$ as follows~\cite{Fan2019prb, evans1982pla}: 
\begin{equation}
\label{equation:Fe}
\mathbf{F}_{i}^{\rm ext} = \mathbf{F}_{\rm e}\cdot\mathbf{W}_{i},
\end{equation}
where $\mathbf{F}_{\rm e}$ is the driving force parameter (with dimensions of inverse length) and was set to \SI{1e-3}{\per\angstrom} in all simulations, and $\mathbf{W}_i$ is the $3 \times 3$ atomic virial tensor of atom $i$ defined as \cite{fan2021neuroevolution} 
\begin{equation}
\label{equation:hnemd}
    \mathbf{W}_i = \sum_{j \neq i} \mathbf{r}_{ij} \otimes  \frac{\partial U_j}{\partial \mathbf{r}_{ji}},
\end{equation}
here $\mathbf{r}_{ij} \equiv \mathbf{r}_j - \mathbf{r}_i$, $\mathbf{r}_i$ is the position of atom $i$, and $U_j$ is the site energy of atom $j$.

Within the framework of linear-response theory, the non-equilibrium ensemble-averaged heat current $\mathbf{J}$ is proportional to the driving-force parameter $\textbf{F}_{\rm e}$:
\begin{equation}
    \langle J^ \alpha \rangle = TV \sum_i \kappa^{\alpha \beta} F_{\rm e} ^ \beta ,
    \label{j_fe}
\end{equation}
where $T$ and $V$ denote the system temperature and volume, respectively, and $\kappa^{\alpha\beta}$ is the thermal conductivity tensor. We compute the instantaneous heat current using the definition \cite{fan2015prb},
\begin{equation}
    \mathbf{J} = \sum_i \mathbf{W}_i \cdot \mathbf{v}_i,
    \label{J}
\end{equation}
where $\mathbf{v}_i$ denotes the velocity of atom $i$.

In the \gls{hnemd} approach, the frequency dependent spectral thermal conductivity $\kappa(\omega)$ is directly accessible through Fourier transformation of the velocity-virial autocorrelation function \cite{Fan2019prb}
\begin{equation}
    \mathbf{K}(t) = \sum_i \langle \mathbf{W}_i (0) \cdot \mathbf{v}_i (t) \rangle.
\end{equation}
The summation runs over all atoms within the system. For simplicity, and because the in-plane thermal transport in graphene based materials is statistically isotropic, we consider only the diagonal components of the thermal conductivity tensor and henceforth drop the tensorial indices, denoting the scalar thermal conductivity simply as $\kappa$ (and similarly $\kappa(\omega)$ for its spectral counterpart). 
The spectral thermal conductivity is obtained via Fourier transform of the virial-velocity autocorrelation function:
\begin{equation}
\label{equation:spectral}
    \kappa(\omega, T) = \frac{2}{VTF_{\rm e}} \int^\infty_{-\infty} \text{d} t e^{i \omega t} K(t),
\end{equation}
where the explicit temperature dependence of the spectral thermal conductivity has been highlighted.

\begin{figure}[h!]
\centering
\includegraphics[width=\columnwidth]{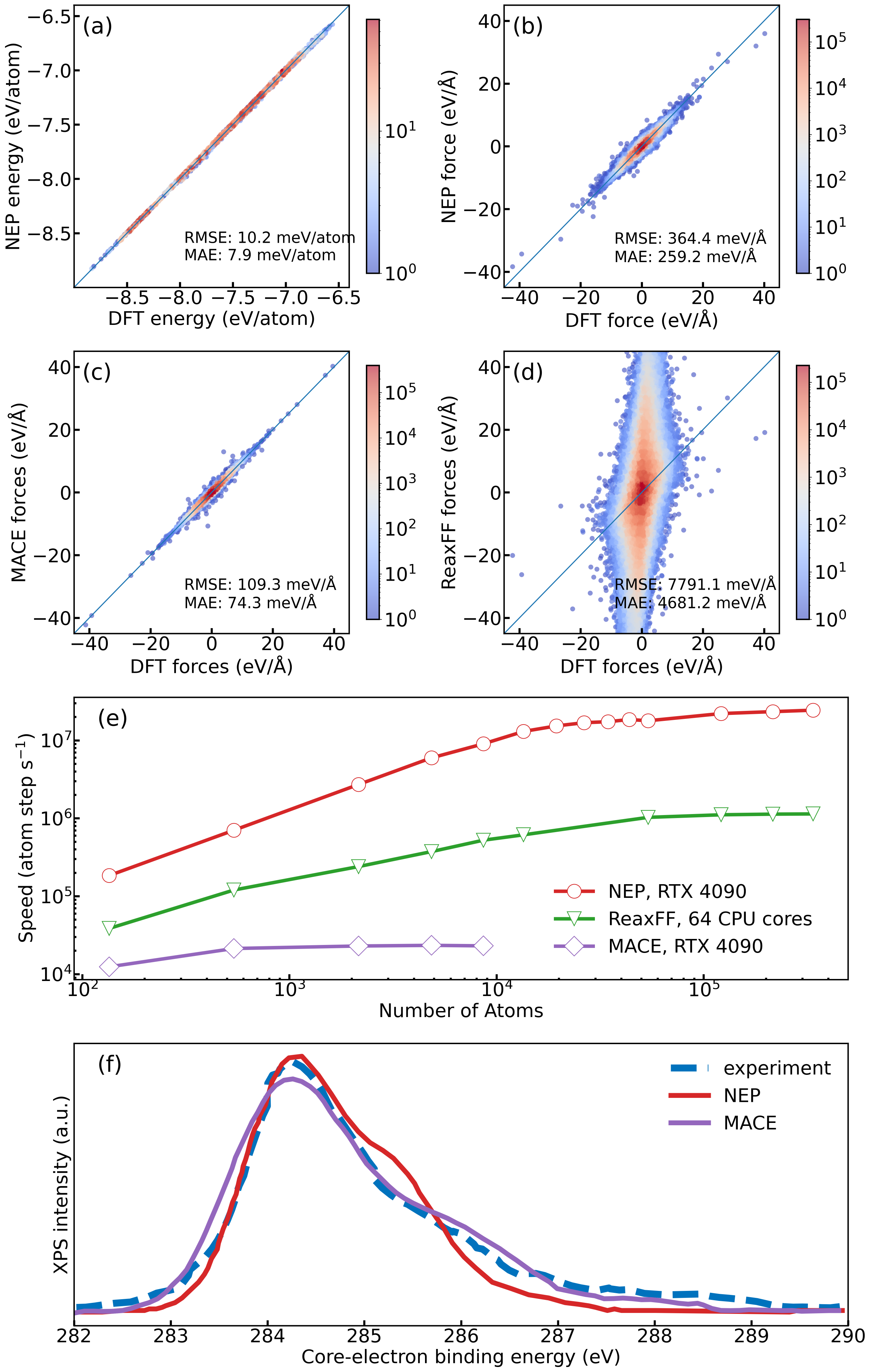}
\caption{Comparison of accuracy and computational speed for \gls{nep}, MACE, and \gls{reaxff} models. (a) Correlation between \gls{dft} reference data and \gls{nep} predictions for total energy across the training set. (b-d) Comparison of predicted forces against \gls{dft} data for (b) \gls{nep}, (c) MACE and (d) \gls{reaxff} across the train dataset. In panels (a)--(d), color intensity represents the local density of data points, and the corresponding \gls{rmse} and \gls{mae} values are provided. (e) Computational speed as a function of the number of atoms in the \gls{go} structure. \gls{nep} and MACE benchmarks were performed on a single RTX 4090 GPU (24 GB memory) using \textsc{gpumd} (version 4.3) and \textsc{ase} (version 3.23.0) packages, respectively; \gls{reaxff} benchmarks were performed on 64 Xeon Platinum 8358P CPU cores (512 GB memory) using \textsc{lammps} package (version 29 Aug 2024)). The MACE and \gls{reaxff} models were obtained from Refs.~\cite{El-Machachi24accelarated} and \cite{kowalik2019atomistic}, respectively. (f) \gls{xps} predictions for the \gls{rgo} structure with an O/C ratio of 0.4 and OH/O ratio of 0.5. The \gls{nep} \gls{xps} profile was calculated for \gls{rgo} annealed at \SI{900}{\kelvin} and equilibrated at \SI{300}{\kelvin}, using \gls{xps} prediction server provided by Ref.~\cite{golze2022accurate}. The \gls{nep} data were shifted horizontally to align with the experimental data from Ref.~\cite{valentini2023tuning}, while the MACE data were extracted from Ref.~\cite{El-Machachi24accelarated}.}
\label{fig:rmse}
\end{figure}

\begin{figure}[htb]
\centering
\includegraphics[width=\columnwidth]{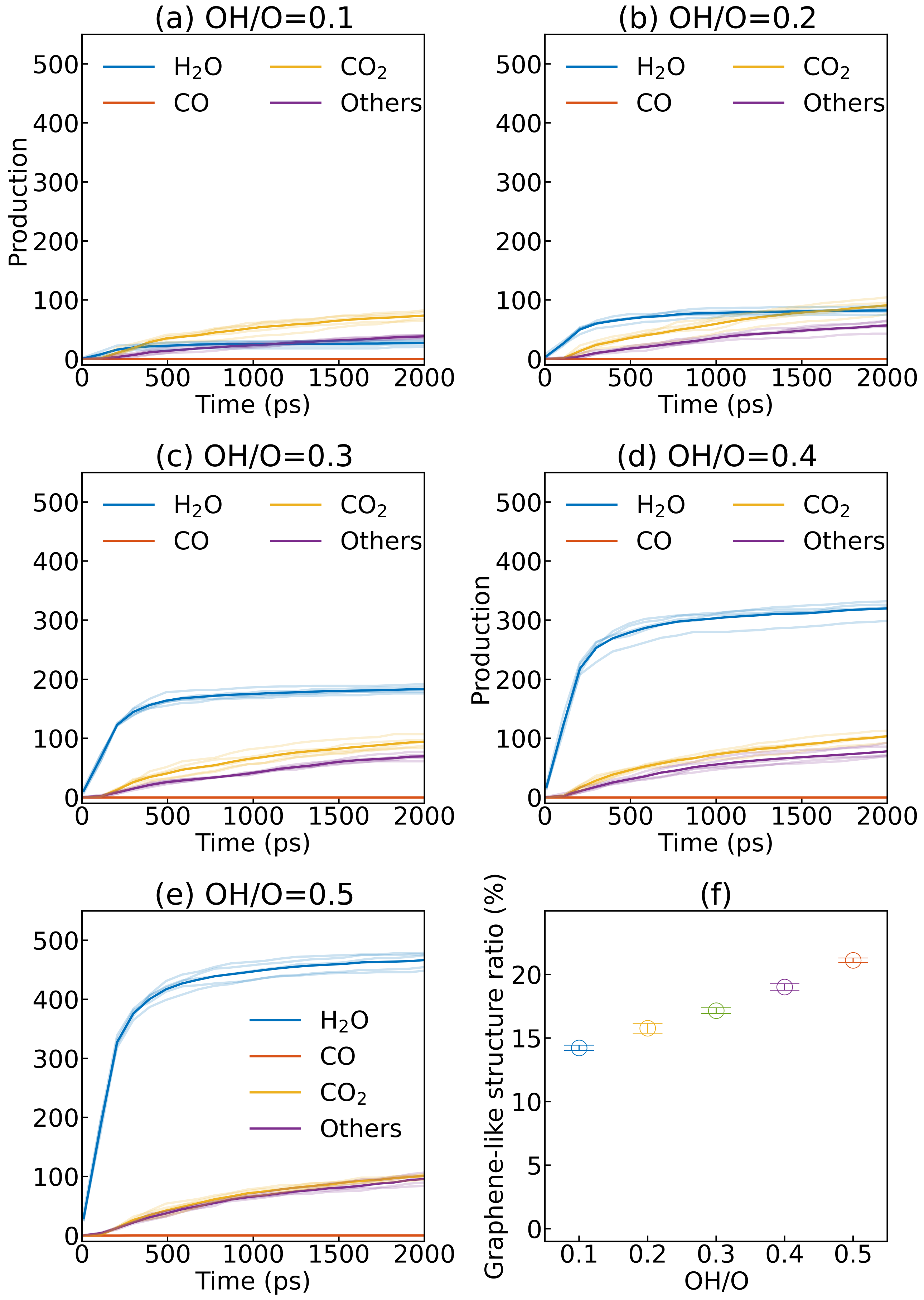}
\caption{Evolution of gaseous byproducts during thermal reduction of \gls{go} at \SI{900}{\kelvin} with a fixed O/C ratio of 0.4. (a)-(e) Time-resolved production of H$_2$O, CO$_2$, CO, and other species during thermal reduction simulations,
for initial OH/O ratios ranging from 0.1 to 0.5: (a) 0.1, (b) 0.2, (c) 0.3, (d) 0.4, and (e) 0.5. Each panel shows the average of five independent simulations (solid lines) with individual trajectories shown in translucence. (f) The ratio of graphene-like structures in the final configurations at \SI{2000}{\pico\second} as a function of initial OH/O ratio.}
\label{fig:product_oh/o}
\end{figure}

\begin{figure}[htb]
\centering
\includegraphics[width=\columnwidth]{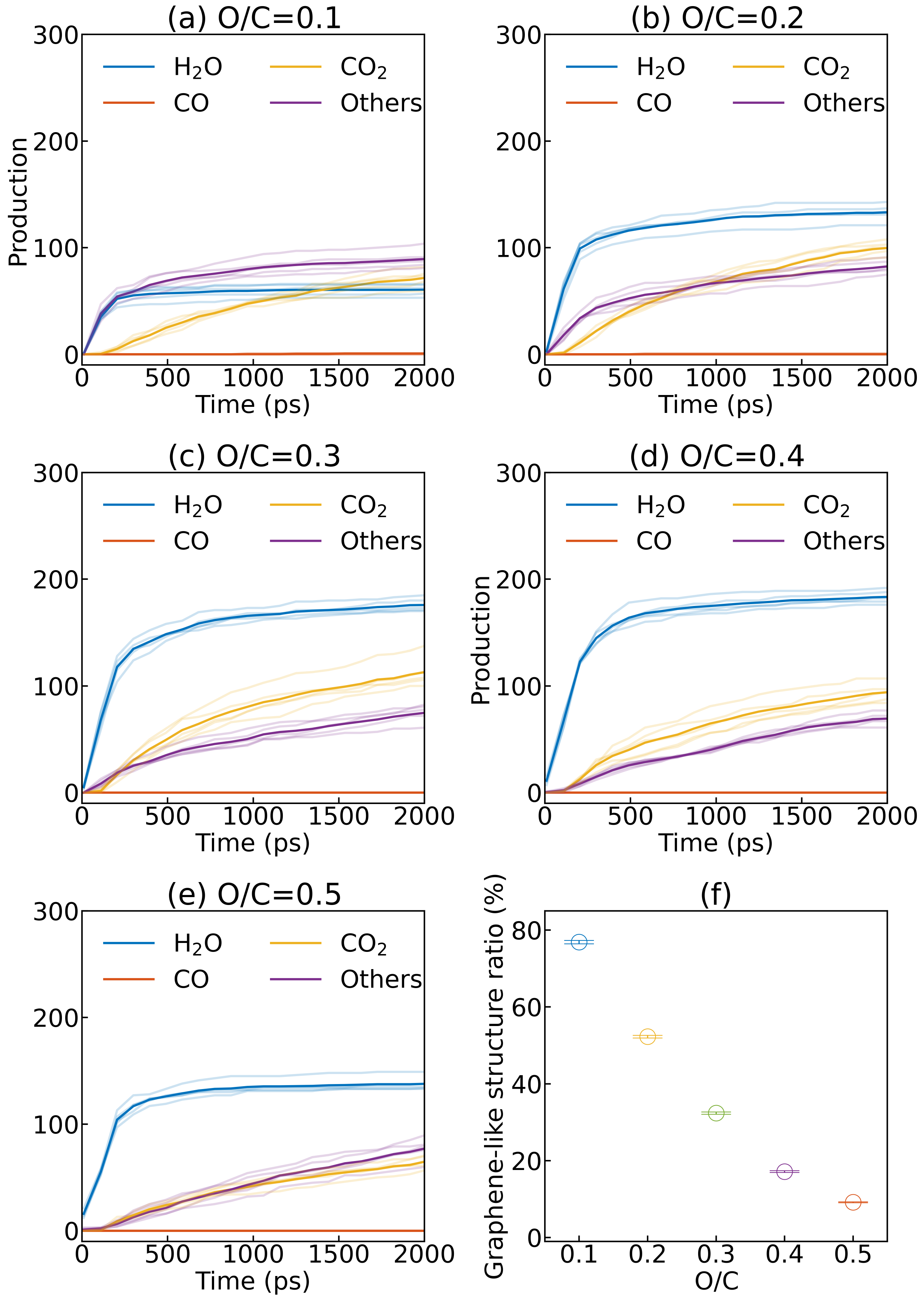}
\caption{Evolution of gaseous byproducts during thermal reduction of \gls{go} at \SI{900}{\kelvin} with a fixed OH/O ratio of 0.3. (a)-(e) Time-resolved production of H$_2$O, CO$_2$, CO, and other species as during thermal reduction simulations,
for initial O/C ratios ranging from 0.1 to 0.5: (a) 0.1, (b) 0.2, (c) 0.3, (d) 0.4, and (e) 0.5. Each panel shows the average of five independent simulations (solid lines) with individual trajectories shown in translucence. (f) The ratio of graphene-like structures in the final configurations at \SI{2000}{\pico\second} as a function of initial O/C ratio.}
\label{fig:product_o/c}
\end{figure}

\begin{figure*}[htb]
\centering
\includegraphics[width=0.8\linewidth]{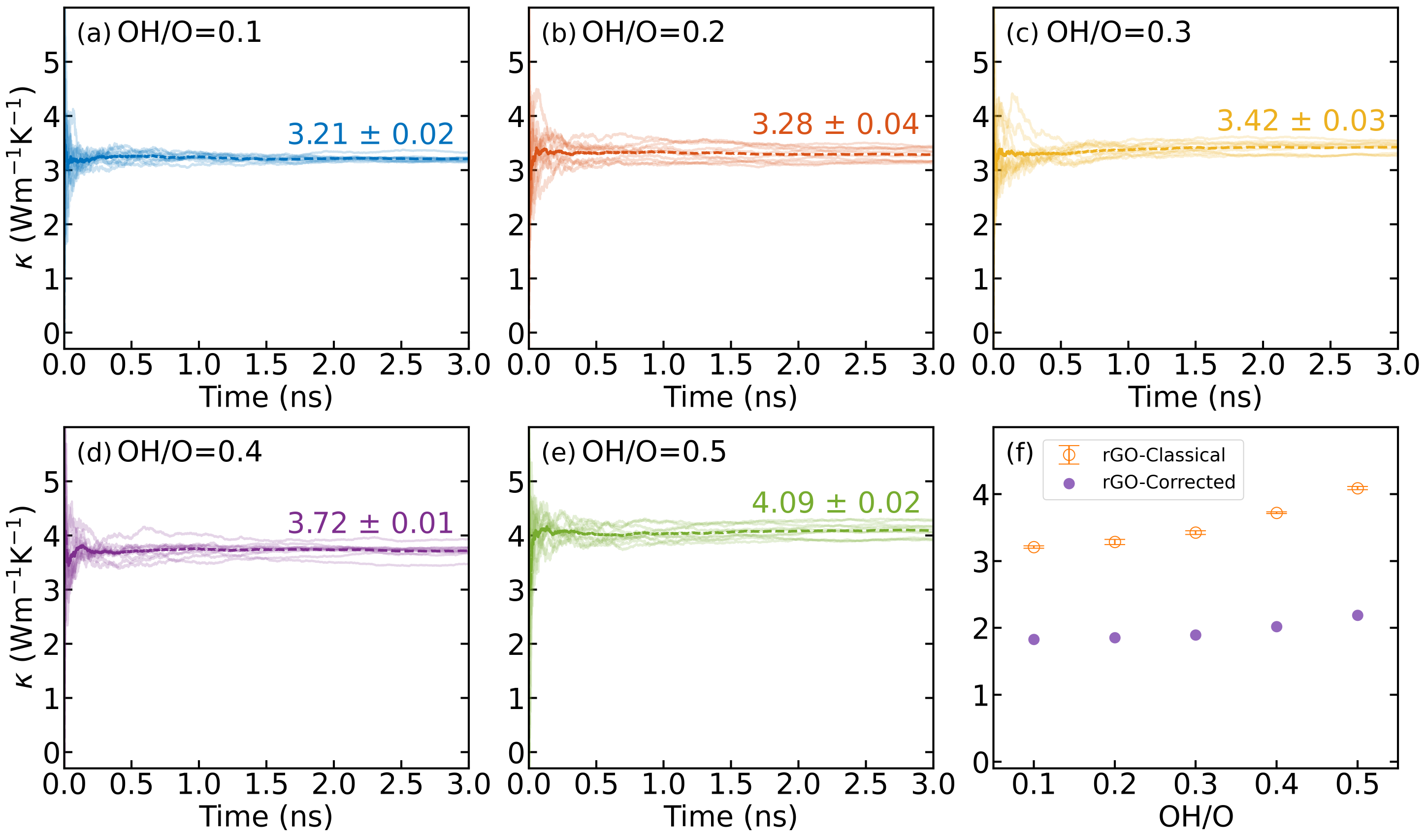}
\caption{Thermal conductivity of \gls{rgo} structures obtained from \gls{hnemd} simulations with a fixed initial O/C ratio of 0.4 and varying initial OH/O ratios. (a–e) Running thermal conductivity for OH/O ratio of (a) 0.1, (b) 0.2, (c) 0.3, (d) 0.4, and (e) 0.5, respectively. In each panel, translucent lines represent ten independent trajectories, while the solid dashed line shows the ensemble mean. The annotated values represent the average thermal conductivity and corresponding standard error calculated at \SI{3}{\nano\second}. (f) Summary comparing the average classical (hollow circles with error bars) and quantum-corrected (solid dots) thermal conductivity as a function of the initio OH/O ratio.}
\label{fig:hnemd_oh/o}
\end{figure*}

\begin{figure*}[htb]
\centering
\includegraphics[width=0.8\linewidth]{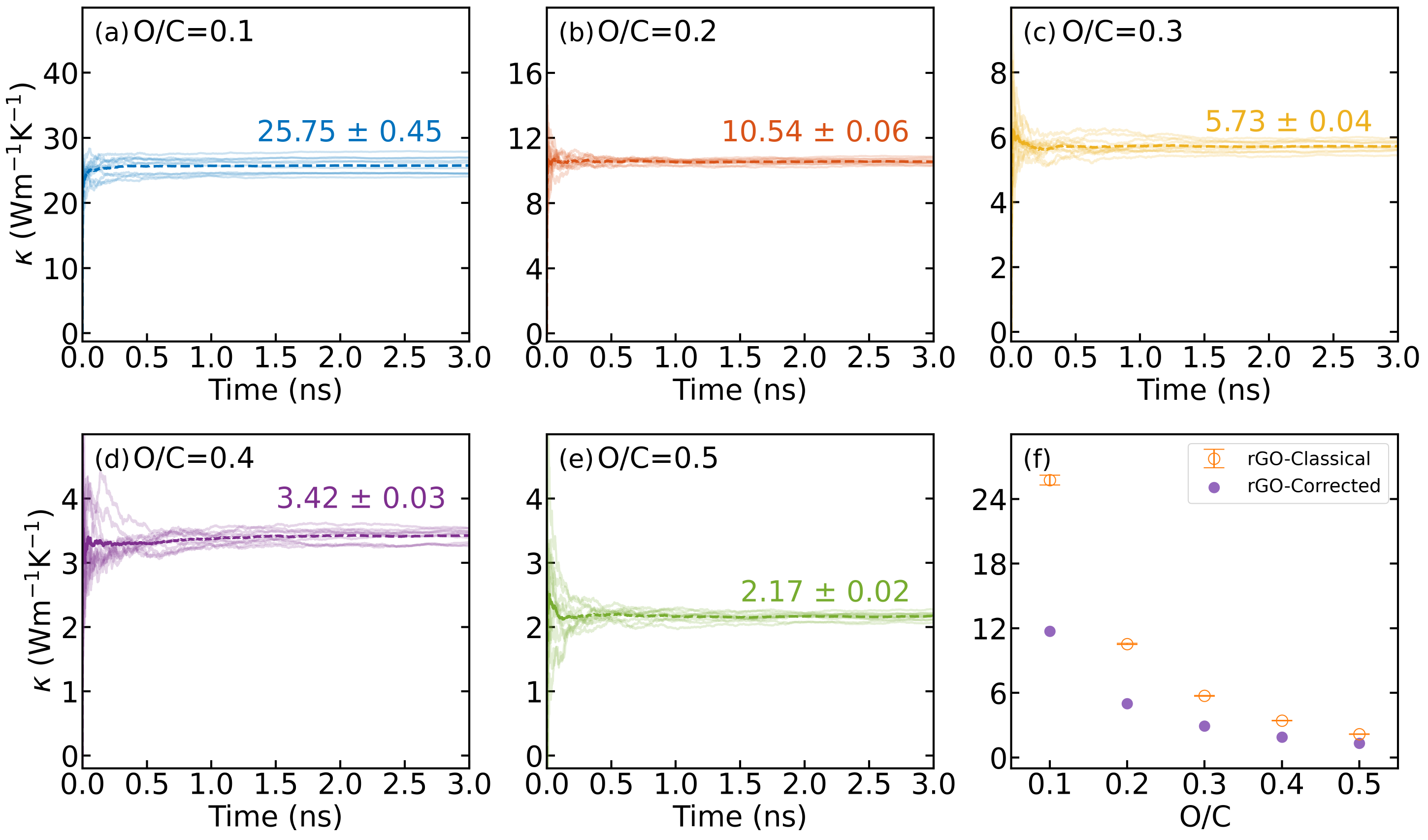}
\caption{Thermal conductivity of \gls{rgo} structures obtained from \gls{hnemd} simulations with a fixed initial OH/O ratio of 0.3 and varying initial O/C ratios. (a–e) Running thermal conductivity for O/C ratio of (a) 0.1, (b) 0.2, (c) 0.3, (d) 0.4, and (e) 0.5, respectively. In each panel, translucent lines represent ten independent trajectories, while the solid dashed line shows the ensemble mean. The annotated values represent the average thermal conductivity and corresponding standard error calculated at \SI{3}{\nano\second}. (f) Summary comparing the average classical (hollow circles with error bars) and quantum-corrected (solid dots) thermal conductivity as a function of the initio O/C ratio.}
\label{fig:hnemd_o/c}
\end{figure*}

\section{Results and Discussion}
The \gls{nep} model developed in this work was trained on the dataset reported in Ref.~\cite{El-Machachi24accelarated}. For comparison, we also evaluated the MACE model trained based on the same reference data~\cite{El-Machachi24accelarated} and the \gls{reaxff} model developed in Ref.~\cite{kowalik2019atomistic}.

As shown in \autoref{fig:rmse}(a), the \gls{nep} model achieves good agreement with the \gls{dft} reference calculations for total energy, with an energy \gls{rmse} of \SI{10.2}{\milli\electronvolt\per\atom}. For atomic forces (\autoref{fig:rmse}(b) and (c)), both \gls{nep} and MACE exhibit excellent agreement with the \gls{dft} reference, with \gls{rmse} values of \SI{364.4}{\milli\electronvolt\per\angstrom} and \SI{109.3}{\milli\electronvolt\per\angstrom}, respectively. In contrast, \gls{reaxff} exhibits significantly larger deviations in force predictions, with a force \gls{rmse} more than one order of magnitude larger (\SI{7791.1}{\milli\electronvolt\per\angstrom}, see \autoref{fig:rmse}(d)), demonstrating considerably lower accuracy for the graphene oxide systems investigated here.

Although the accuracy of our \gls{nep} model is lower than that of MACE, which benefits from message-passing architectures and equivariant feature representations within its graph neural network framework~\cite{batatia2022mace, kovacs2023evaluation}, the primary reason for selecting \gls{nep} is its superior computational efficiency. This efficiency is crucial for thermal conductivity calculations, which require extensive sampling over long timescales and large system sizes. 

To demonstrate this advantage, we performed a benchmark timing test using an initial GO structure generated with the parameter set $\mathbf{P} = [0.4, 0.5, 0]$ (containing 135 atoms). This unit cell was then replicated into supercells of increasing sizes. NVT-ensemble \gls{md} simulations were run for 2000 steps at \SI{300}{\kelvin} using \gls{nep}, MACE, and \gls{reaxff}. As shown in \autoref{fig:rmse} (e), the computational speed for all models initially increases with system size and reaches saturation for relatively larger systems.

Using a single RTX 4090 GPU, \gls{nep} implemented in the \textsc{GPUMD}  package achieves a speed exceeding \SI{1e7}{\atomstepsecond} for systems containing hundreds of thousands of atoms. In contrast, MACE implemented in the \textsc{ASE} package~\cite{hjorth2017the} is practical only for systems up to thousands of atoms, exhibiting a computational speed three orders of magnitude lower,  near \SI{1e4}{\atomstepsecond}. Furthermore, compared with the empirical \gls{reaxff} model implemented in the \textsc{LAMMPS} package~\cite{LAMMPS} running on 64 CPU cores, our machine-learned \gls{nep} model demonstrates a computational speed that is more than one order of magnitude higher. These results justify the choice of the \gls{nep} model, as it offers the necessary balance of accuracy and efficiency required for the large-scale, long-timescale \gls{md} simulations in this study. We note that although the accuracy of \gls{nep} is relatively lower than that of MACE, \gls{nep} predicted a reasonable XPS profile for rGO after thermal reduction simulations (\autoref{fig:rmse}(f)), which aligns well with experimental data~\cite{valentini2023tuning} and MACE prediction~\cite{El-Machachi24accelarated}, demonstrating the reliability of our developed NEP model.

To elucidate the structural evolution during the thermal reduction of \gls{go} to \gls{rgo}, we monitored the production of the primary gaseous byproducts through two complementary series of simulations: first by keeping the O/C ratio fixed while systematically varying the OH/O ratio (\autoref{fig:product_oh/o}), and second by keeping the OH/O ratio fixed while systematically varying the O/C ratio (\autoref{fig:product_o/c}). To ensure statistical reliability, we performed five independent thermal reduction simulations for each set of OH/O and O/C ratios. We quantified the fraction of "graphene-like" atoms in the resulting \gls{rgo} structure using \gls{ptm}~\cite{Larsen2016robust} as implemented in \textsc{OVITO}~\cite{stukowski2009visualization}，applying a \gls{rmsd} cutoff of 0.15.

As shown in \autoref{fig:product_oh/o} (a)-(e), where the O/C ratio is fixed at 0.4 and the OH/O ratio ranges from 0.1 to 0.5, the major desorption products are $\rm H_2O$ (blue line), $\rm CO_2$ (yellow line), CO (pink line), and a small fraction of other species (purple line). We further observe that increasing the initial OH/O ratio leads to a substantial rise in the total yield of gaseous byproducts. Among these, H$_2$O production is most strongly influenced, exhibiting the sharpest increase with higher hydroxyl content. In contrast, while CO evolution was monitored throughout the simulations, its final yield remains negligible across all studied ratios. \autoref{fig:product_oh/o}(f) reports the structural quality of the final reduced frames.  As the initial OH/O ratio increases, the recovery of the graphene lattice, evidenced by the fraction of \gls{ptm}-identified hexagonal environments, exhibits a noticeable improvement. Specifically, it rises from $\sim 14\%$ at OH/O = 0.1 to  $\sim 21\%$ at OH/O = 0.5, representing an approximate 50\% relative increase. This structural restoration is attributed to the preferential desorption of oxygen as H$_2$O at higher hydroxyl concentrations, a non-destructive pathway that preserves the underlying carbon skeleton, whereas lower OH/O ratios favor carbon-consuming pathways, with CO$_2$ evolution acting as the dominant source of carbon loss, which generates lattice vacancies.

In a complementary analysis, we fixed the OH/O ratio at 0.3 while systematically varying the O/C ratio from 0.1 to 0.5 to investigate the influence of overall oxygen content on the thermal reduction process (see \autoref{fig:product_o/c}). As illustrated in the gas evolution profiles (\autoref{fig:product_o/c}(a)-(e)), unlike the OH/O variation, changes in the O/C ratio do not drive a continuous increase in H$_2$O evolution; instead, H$_2$O production saturates and even exhibits a moderate decrease at high O/C ratios (0.4–0.5). Crucially, while CO$_2$ production plateaus between O/C = 0.3 and 0.4, there is a marked rise in the evolution of 'other' gaseous byproducts for O/C=0.5 (see purple line in \autoref{fig:product_o/c}(e)). This indicates that at the highest oxidation levels, carbon etching is increasingly driven by these alternative decomposition pathways alongside CO$_2$. Most notably, the \gls{ptm} analysis (\autoref{fig:product_o/c}(f)) reveals a drastic degradation in structural quality: the fraction of graphene-like carbon significantly drops from ~77\% at O/C = 0.1 to just ~9\% at O/C = 0.5. This confirms that higher initial oxygen content activates aggressive carbon-consuming reaction channels, shifting from simple CO$_2$ evolution to complex fragmentation, severely hindering the restoration of the pristine graphitic network.

\begin{figure}[htb]
\centering
\includegraphics[width=\columnwidth]{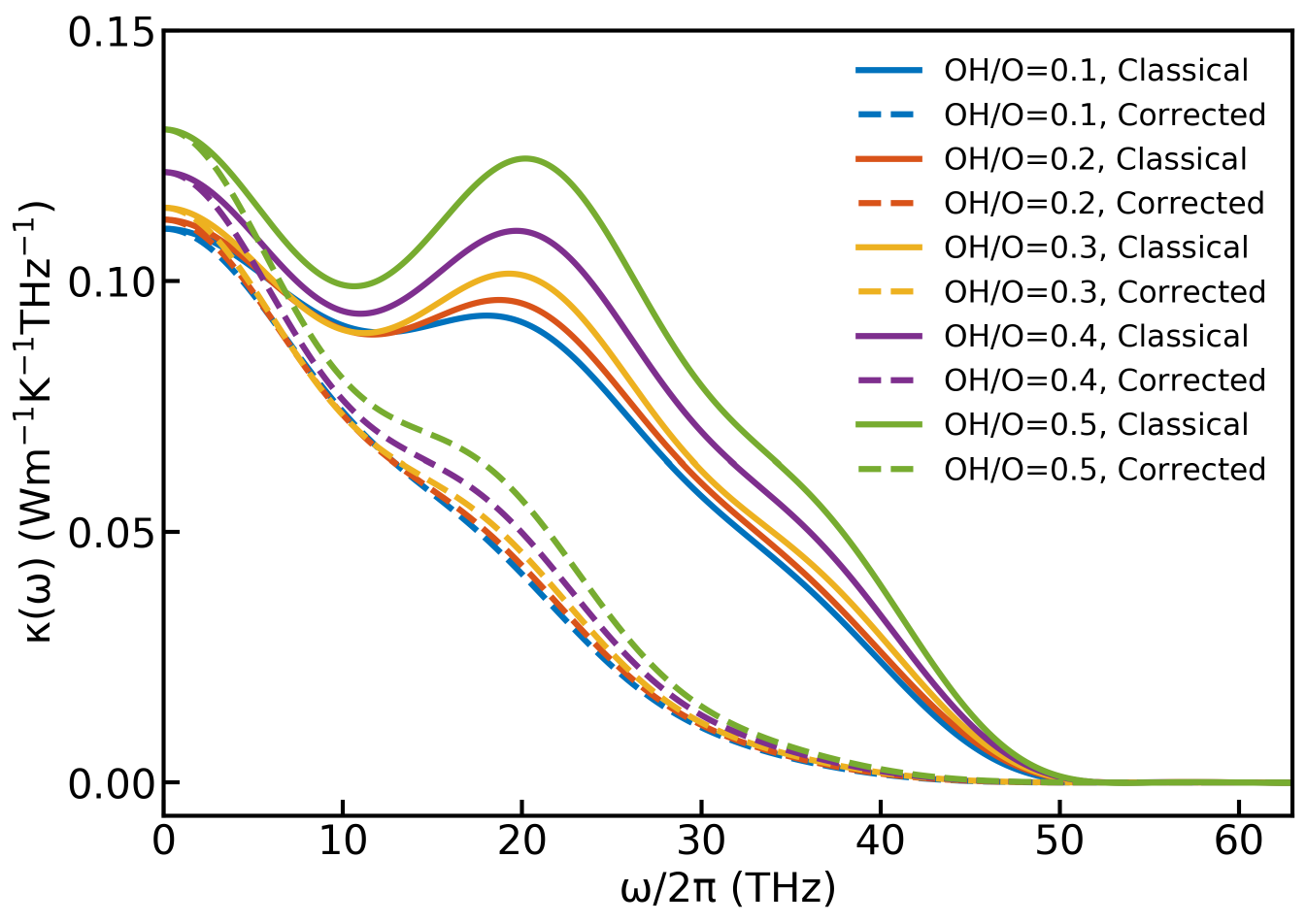}
\caption{Classical (solid lines) and quantum-corrected (dash lines) spectral thermal conductivity, $\kappa(\omega)$, as a function of phonon frequency for \gls{rgo} structures with a fixed O/C ratio of 0.4 and varying OH/O ratios from 0.1 to 0.5. Each curve denotes the average of ten independent runs (two orthogonal in-plane directions across five separately generated structures) to ensure statistical robustness.}
\label{fig:shc_oh/o}
\end{figure}

\begin{figure}[htb]
\centering
\includegraphics[width=\columnwidth]{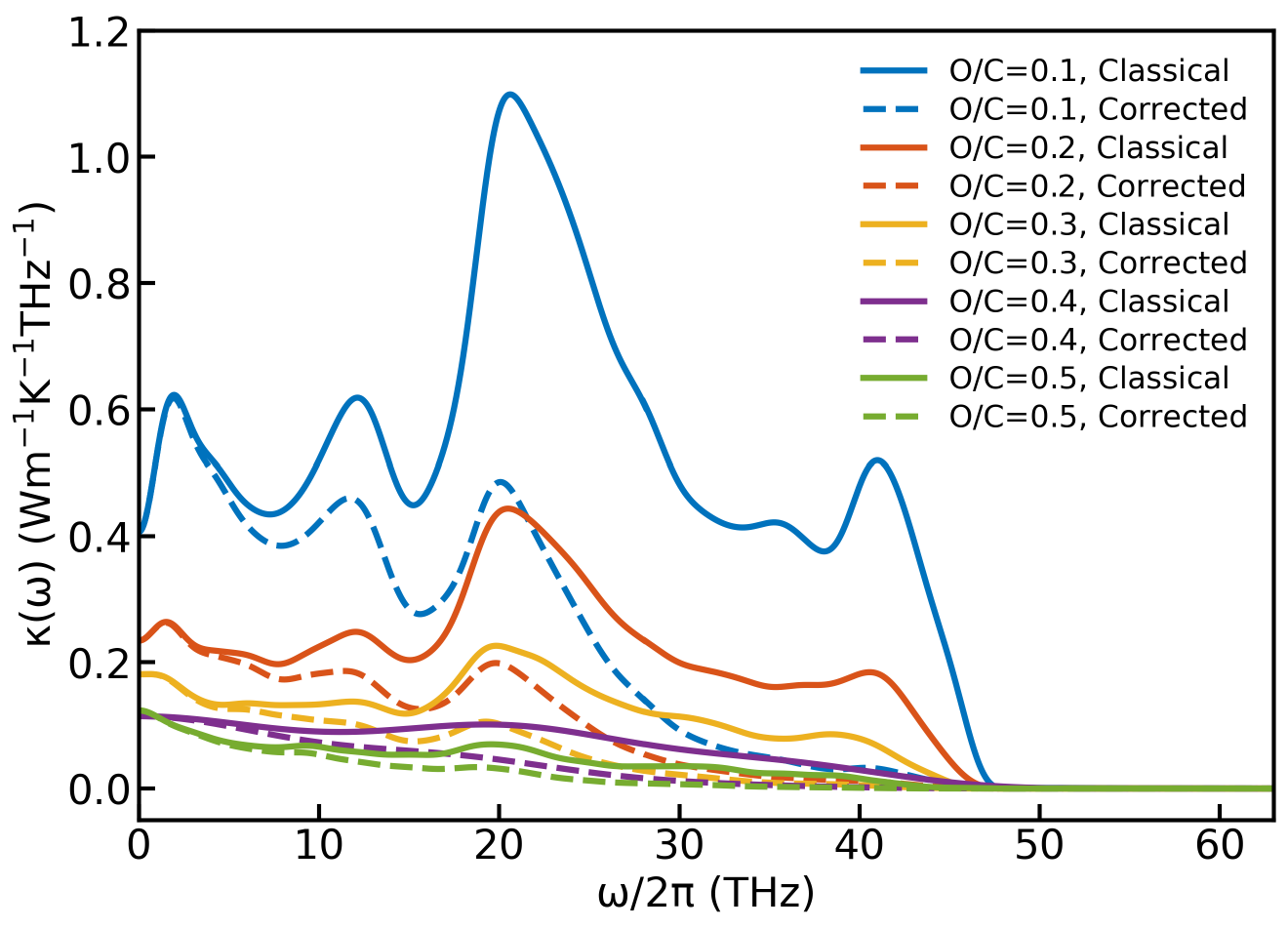}
\caption{Classical (solid lines) and quantum-corrected (dash lines) spectral thermal conductivity, $\kappa(\omega)$, as a function of phonon frequency for \gls{rgo} structures with a fixed OH/O ratio of 0.3 and varying O/C ratios from 0.1 to 0.5. Each curve denotes the average of ten independent runs (two orthogonal in-plane directions across five separately generated structures) to ensure statistical robustness.}
\label{fig:shc_o/c}
\end{figure}

\begin{figure}[htb]
\centering
\includegraphics[width=\columnwidth]{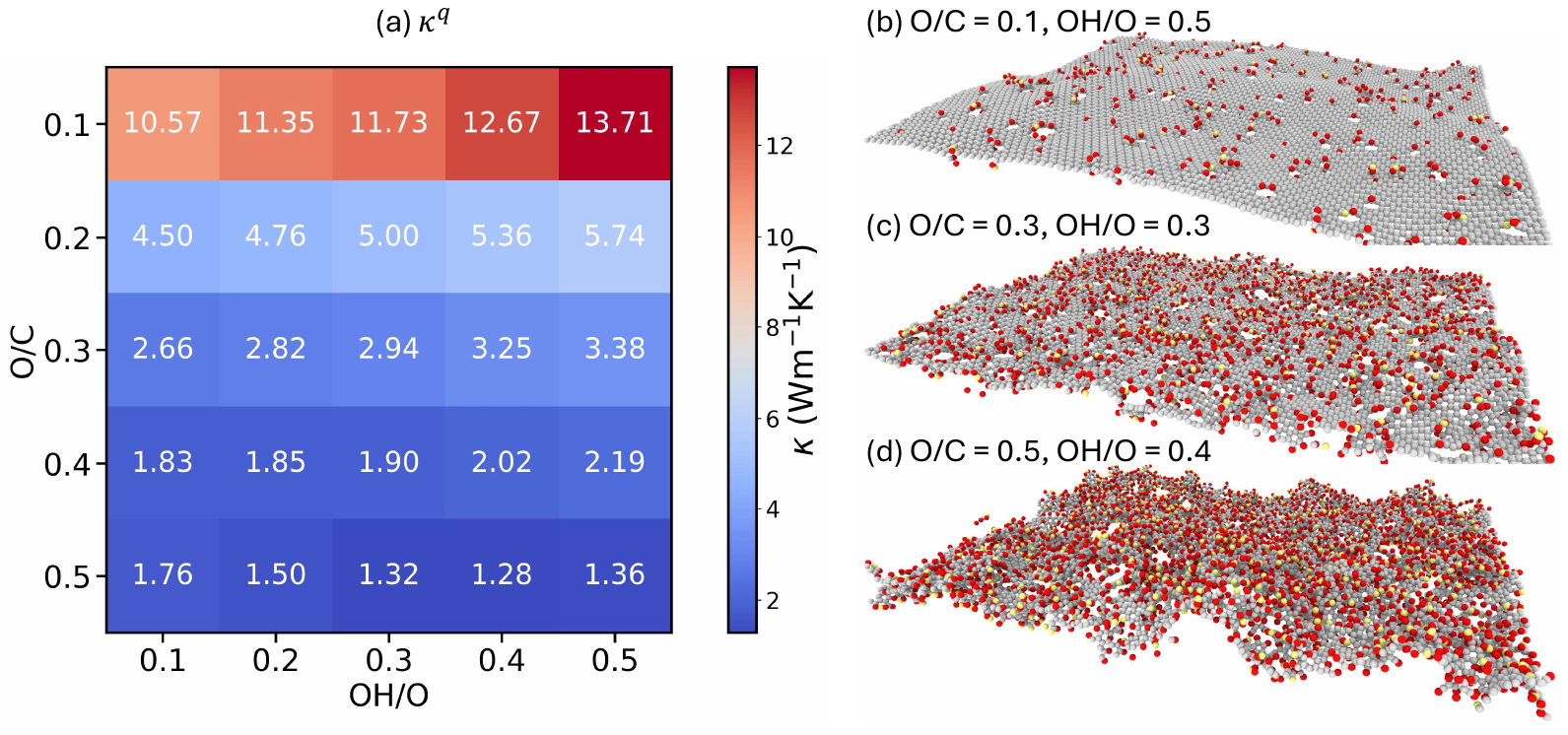}
\caption{(a) Computed quantum-corrected thermal conductivity ($\kappa^q$) of \gls{rgo} structures across the full range of investigated compositional ratios. The heatmap displays the ensemble-averaged conductivity as a function of the initial O/C and OH/O ratios. For each grid point, the reported value represents the mean obtained from ten independent \gls{hnemd} simulations (comprising heat transport calculations along two orthogonal in-plane directions for five independent structural models generated from separate thermal reduction trajectories). (b-d) Three representative atomic snapshots for \gls{rgo} showing the (b) highest $\kappa^q$ (O/C = 0.1, OH/O = 0.5), (c) intermediate $\kappa^q$ (O/C = 0.3, OH/O = 0.3), and (d) lowest $\kappa^q$ (O/C = 0.5, OH/O = 0.4). Visualizations were created using \textsc{ovito} package~\cite{stukowski2009visualization}, with carbon, oxygen, and hydrogen atoms colored gray, red, and yellow, respectively.}
\label{fig:imshow}
\end{figure}

We computed the thermal conductivity of the resulting \gls{rgo} structures using the \gls{hnemd} simulations based on \autoref{equation:hnemd} (see \autoref{fig:hnemd_oh/o} and \autoref{fig:hnemd_o/c}). For each set of O/C and OH/O ratios, we performed ten independent simulations (two lateral directions for each of the five configurations from independent thermal reduction simulations) to ensure statistical robustness. As the running thermal conductivity stabilizes at \SI{3}{\nano\second} (\autoref{fig:hnemd_oh/o}(a)-(e) and \autoref{fig:product_o/c}(a)-(e));  production values were calculated from this point, with error bars representing the standard error of the mean.

In \autoref{fig:hnemd_oh/o}(f), for the series with a fixed initial O/C ratio of 0.4,  we observe that increasing the initial OH/O ratio leads to a moderate enhancement in the classical thermal conductivity of the \gls{rgo} structures. Specifically, the conductivity rises from \SI{3.21\pm0.02}{\watt\per\meter\per\kelvin} at OH/O = 0.1 to \SI{4.09\pm0.02}{\watt\per\meter\per\kelvin} at OH/O = 0.5, representing an $\sim 27\%$ increase. In sharp contrast, a substantial decrease in the classical thermal conductivity of the \gls{rgo} structures was observed at a fixed OH/O ratio of 0.3 as the O/C ratio increased from 0.1 to 0.5. This results in an order--of--magnitude reduction, dropping from \SI{25.73\pm0.46}{\watt\per\meter\per\kelvin} at O/C = 0.1 to \SI{2.16\pm0.02}{\watt\per\meter\per\kelvin} at OH/O = 0.5.

We attribute the enhancement in thermal conductivity at higher OH/O and lower O/C ratios to two synergistic factors: structural restoration and mechanical stiffening. First, the substantial rise in the graphene-like carbon fraction (\autoref{fig:product_oh/o}(f) and \autoref{fig:product_o/c}(f)) creates more continuous and interconnected transport pathways, effectively suppressing phonon scattering and damping. Second, this structural recovery at higher OH/O and lower O/C ratios increases the material's Young's modulus~\cite{el2025mechanical}; this stiffening is directly linked to higher phonon group velocities, which further accelerate heat propagation through the lattice. Notably, we find that the thermal conductivity of \gls{rgo} is much more sensitive to O/C ratio than OH/O ratio, because low O/C ratios significantly suppress carbon-consuming gaseous byproducts (\autoref{fig:product_o/c}(a)), thereby preserving the carbon skeleton and leading to a much stronger dependence of the graphene-like structure fraction on the initial oxygen content.

Given the strong structural disorder in \gls{rgo} and this system has high vibrational frequencies, accurate thermal transport predictions require accounting for nuclear quantum effects. While path-integral techniques~\cite{ying2025highly} capture these effects, it is currently incompatible with thermal conductivity calculations due to the nonlinearity of the heat current operator~\cite{zeng2025lattice}. Therefore, we adopted a feasible quantum-correction method based on the harmonic approximation, which has been rigorously validated for disordered materials~\cite{wang2023prb, liang2023prb, wang2025density} and even liquid water\cite{xu2023jcp, xu2025nep}. This is achieved by multiplying it by the ratio between quantum and classical modal heat capacities~\cite{fan2017nl, k2016vibrational, Lv2016direct}:
\begin{equation}
\label{equation:spectral_correct}
    \kappa^q(\omega) =\kappa(\omega) \frac{x^2e^x}{(e^x -1)^2},
\end{equation}
where $\kappa^q(\omega)$ and $\kappa(\omega)$ are classical and quantum-corrected spectral thermal conductivity, respectively, and $x = \hbar \omega / k_B T$, with $\hbar$ the reduced Planck constant and $k_B$ the Boltzmann constant. We emphasize that this quantum-correction approach is valid primarily in disordered systems, where vibrational modes exhibit short lifetimes and mean free paths. In such cases, the population effects on elastic scattering processes remain negligible, justifying the simple multiplicative heat-capacity correction.

To implement this correction, we decomposed the total thermal conductivity $\kappa$ from  \gls{hnemd} calculations performed at \SI{300}{\kelvin} into its frequency-resolved spectral counterpart $\kappa(\omega)$ using \autoref{equation:spectral}. \autoref{fig:shc_oh/o} and \autoref{fig:shc_o/c} present both the classical and quantum-corrected spectral thermal conductivity curves for each investigated OH/O and O/C ratio. Upon integrating $\kappa^q(\omega)$ for all ratios of \gls{rgo} at \SI{300}{\kelvin} in \autoref{fig:shc_o/c} and \autoref{fig:shc_oh/o}, we present the quantum-corrected $\kappa$ in \autoref{fig:hnemd_oh/o} (f) and \autoref{fig:hnemd_o/c} (f) with solid circles. For the OH/O series (\autoref{fig:hnemd_oh/o}), the $\kappa(\omega)$ profile exhibits remarkable similarity across all ratios, consistently featuring a single dominant peak at $\sim$\SI{20}{\tera\hertz}. This similarity persists after quantum correction, which simply scales down the high-frequency contributions. In contrast, varying the O/C ratio (\autoref{fig:hnemd_o/c}) significantly alters the spectral features. Unlike the uniform profiles observed in the OH/O series, the low O/C structures (0.1 and 0.2) display a complex multi-peak structure with distinct contributions at $\sim$\SI{2}{\tera\hertz}, $\sim$\SI{12}{\tera\hertz}, $\sim$\SI{20}{\tera\hertz}, and $\sim$\SI{41}{\tera\hertz}. However, as the O/C ratio increases, these features are progressively suppressed, resulting in a featureless profile at high oxidation levels. The quantum correction further dampens these high-frequency modes. Finally, compared with the classical $\kappa$, the quantum-corrected $\kappa^q$ are reduced by $\sim$45\% to $\sim$60\% across all structures.

\begin{table}[htb]
\centering
\setlength{\tabcolsep}{2.5mm}
\caption{Comparison of thermal conductivity $\kappa$ (W m$^{-1}$ K$^{-1}$) of monolayer \gls{2d} materials, including \gls{go}, \gls{rgo}, graphene, qHP (quasi-hexagonal phase) and qTP (quasi-tetragonal phase) fullerene systems, phosphorene (P), MoS$_2$, and $h$-BN. For the \gls{rgo}, \gls{go}, and graphene structures calculated in this work, the reported values for \gls{rgo} and \gls{go} include quantum corrections. Predictions obtained via \gls{hnemd}, NEMD (nonequilibrium molecular dynamics), EMD (equilibrium molecular dynamics), and BTE (Boltzmann transport equation) methods as well as experimental measurements, are included for comparison.}
\label{table:k}
\begin{tabular}{llllll} %
\hline
Monolayer \gls{2d} materials & Method & $\kappa$ \\
\hline
\gls{rgo} (This work) & HMEMD & 1.28-- 13.71  \\
\gls{rgo} (3--5 nm thickness)~\cite{schwamb2009electrical} & Experiment & 0.14-- 2.87  \\
\gls{rgo} (1 nm thickness)~\cite{mahanta2012thermal}& Experiment &  18(2) \\
\gls{go} (This work) & HNEMD &  2.89 -- 13.98 \\
\gls{go} (O/C $\leq$ 0.04)~\cite{zhang2014tailoring} & NEMD &  $\sim$40 -- $\sim$420 \\
\gls{go} (O/C $\leq$ 0.2)~\cite{lin2014thermal} & NEMD & 18--75 \\
\gls{go} (0.05 $\leq$ O/C $\leq$ 0.2)~\cite{mu2014thermal} & NEMD & 8.8--28.8 \\
graphene (This work) & HNEMD & 1391(43) \\
graphene~\cite{xu2014length} & Experiment & 1689--1813 \\
graphene~\cite{Wu2024JChemPhys} &HNEMD &1855(56) \\
graphene~\cite{fan2017prb} & EMD & 2900(100) \\
graphene~\cite{ying2022thermal} & HNEMD & 2807.3\\
graphene~\cite{ying2022thermal} & EMD &  3067.4\\
blue phosphorene ~\cite{ying2023variable}& HNEMD & 128(3)  \\
violet P ~\cite{ying2023variable}& HNEMD & 2.36(0.05)  \\
black P (zigzag) ~\cite{ying2023variable}& HNEMD & 78.4(0.4)  \\
black P (armchair) ~\cite{ying2023variable}& HNEMD & 12.5(0.2)  \\
MoS$_2$ ~\cite{jiang2025accurate} & HNEMD & 150.4 (6.2)\\
MoS$_2$ ~\cite{yang2020experimental} & Experiment & 86, 100 \\
qHP-$x$ C$_{24}$~\cite{li2025anisotropic}& HNEMD  & 233(5)  \\
qHP-$y$ C$_{24}$~\cite{li2025anisotropic}& HNEMD  & 341(9)  \\
qTP-C$_{24}$~\cite{li2025anisotropic}& HNEMD  & 272(9)   \\
qHP-$x$ C$_{60}$~\cite{dong2023ijhmt}& HNEMD  & 102(3)  \\
qHP-$y$ C$_{60}$~\cite{dong2023ijhmt}& HNEMD  & 107(7)  \\
$h$-BN~\cite{tan2025coherent} & HNEMD &  544(10) \\
$h$-BN~\cite{qiran2019high} & Experiment & 751 \\
$\alpha$-graphyne~\cite{Yang2018pccp} & BTE & 21.11   \\
biphenylene (armchair)~\cite{Veeravenkata2021Carbon} & BTE &  166\\
biphenylene (zigzag)~\cite{Veeravenkata2021Carbon} & BTE &  254\\
biphenylene (armchair)~\cite{ying2022thermal} & HNEMD &  213.1\\
biphenylene (zigzag)~\cite{ying2022thermal} & HNEMD &  203.5\\
biphenylene (armchair)~\cite{ying2022thermal} & EMD &  232.3\\
biphenylene (zigzag)~\cite{ying2022thermal} & EMD &  226.3\\
\hline
\label{table:kappa_compare}
\end{tabular}
\end{table}

To provide a comprehensive overview, we computed the quantum-corrected thermal conductivity, $\kappa^q$, for the full range of initial O/C and OH/O ratios, as summarized in the heatmap in \autoref{fig:imshow}. The results confirm that $\kappa^q$ is predominantly governed by the O/C ratio, decreasing by an order of magnitude as the initial oxidation level increases from 0.1 to 0.5. In contrast, increasing the OH/O ratio consistently yields a moderate enhancement in thermal conductivity for O/C ratio below 0.5, reflecting the improved recovery of the graphene lattice discussed earlier. However, a notable exception is observed at the highest oxidation level (O/C = 0.5), where $\kappa^q$ slightly decreases with increasing OH/O ratio. This inversion suggests that at saturation coverage, the high density of hydroxyl groups may trigger aggressive decomposition pathways or defect clustering, likely analogous to oxidative etching~\cite{liu2008graphene}, that outweighs the lattice-healing effects observed at lower oxidation levels (see \autoref{fig:imshow}(b-d)). Consequently, we observe a maximum \gls{rgo} thermal conductivity of \SI{13.71}{\watt\per\meter\per\kelvin} (\autoref{fig:imshow}(b); at O/C = 0.1, OH/O = 0.5), and a minimum of \SI{1.28}{\watt\per\meter\per\kelvin} (\autoref{fig:imshow}(d); at O/C = 0.5, OH/O = 0.4). 

Finally, \autoref{table:kappa_compare} benchmarks our calculated $\kappa^q$ values against other monolayer \gls{2d} materials. The predicted thermal conductivity of \gls{rgo} (\SIrange{1.28}{13.71}{\watt\per\meter\per\kelvin}) falls within the range of experimental measurements for samples with similar carbon ratios~\cite{schwamb2009electrical, mahanta2012thermal}, which is orders of magnitude lower than that of pristine graphene ($>$\SI{1000}{\watt\per\meter\per\kelvin}) and hexagonal boron nitride, but is comparable to violet phosphorene and black phosphorene, highlighting its potential for thermoelectric applications~\cite{li2018thermoelectric} where low thermal conductivity is advantageous. 

Our simulations estimate that the thermal conductivity of \gls{rgo} is comparable to or even lower than that of the initial \gls{go} models (\SIrange{4.79}{30.00}{\watt\per\meter\per\kelvin}, lowest at OH/O=O/C=0.5, highest at OH/O=O/C=0.1), a trend that contrasts with previous experimental measurements where conductivity typically increases after reduction~\cite{renteria2015strongly}. We attribute this discrepancy to two factors. First, our initial \gls{go} models were constructed on ideal graphene lattices without atomic vacancies; consequently, the thermal reduction process, which drives carbon loss and vacancy formation, effectively degrades the phonon transport relative to the pristine starting state. Second, experimental \gls{go} and \gls{rgo} are typically multi-layer systems, where high-temperature annealing facilitates defect healing across layers~\cite{ying2025chemifriction}, a recovery mechanism not present in our monolayer simulations. We leave the investigation of these effects for future work; our present study therefore serves as a foundation for further understanding thermal transport in multi-layer and bulk \gls{rgo} and \gls{go} systems.

\section{Summary and conclusions}
In summary, we have established a computationally efficient and predictive atomistic framework to investigate the complex interplay between chemical reduction and thermal transport in \gls{go}. By developing a neuroevolution potential (\gls{nep}--\gls{go}) trained on \gls{dft} data, we achieved a computational speedup of several orders of magnitude compared to MACE and \gls{reaxff} models while maintaining a reasonable accuracy. This capability enabled large-scale, long-timescale \gls{md} simulations of thermal reduction and heat transport that were previously computationally prohibitive.

Our simulations reveal that the thermal conductivity of \gls{rgo} is not solely governed by the degree of oxidation alone but is critically dependent on the specific chemical composition of the initial state. We identified two distinct structural evolution pathways: (i) generally, increasing the initial OH/O ratio promotes non-destructive water desorption, leading to substantial lattice recovery and enhanced thermal transport, though this trend inverts at the highest oxidation level (O/C = 0.5) where hydroxyl saturation triggers aggressive lattice disruption rather than healing; and (ii) increasing the initial O/C ratio activates aggressive carbon-etching pathways via CO and CO$_2$ evolution, resulting in severe defect formation and suppressed thermal conductivity. Furthermore, by decomposing the thermal conductivity into its spectral components and applying quantum-statistical corrections, we demonstrated that nuclear quantum effects significantly suppress high-frequency vibrational modes in these disordered systems, reducing the predicted thermal conductivity by approximately 50\% compared to classical predictions.

Quantitatively, the quantum-corrected thermal conductivity of monolayer \gls{rgo} spans a range of \SIrange{1.28}{13.71}{\watt\per\meter\per\kelvin}. While these values are orders of magnitude lower than pristine graphene, they render \gls{rgo} a promising candidate for thermoelectric applications where low lattice thermal conductivity is desirable. By demonstrating how the specific ratios of oxygen functional groups can be tuned to design target thermal conductivities, we pave the way for future investigations that extend these principles to topological structures and multi-layer bulk systems, ultimately enabling precise defect-engineering strategies for advanced thermal management.

\vspace{0.5cm}

\begin{acknowledgments}
This work is supported by the National Science and Technology Advanced Materials Major Program of China (No. 2024ZD0606900). 
ZF is supported by the Science Foundation from Education Department of Liaoning Province
(No. LJ232510167001).
Jinglei Yang and the HKUST personnel are supported by the Project of Hetao Shenzhen-Hong Kong Science and Technology Innovation Cooperation Zone (HZQB-KCZYB-2020083) and the Innovation and Technology Commission of Hong Kong (ITC-CNERC14SC01).
\end{acknowledgments}

\vspace{0.5cm}

\noindent{\textbf{Data availability:}}

Complete input and output files for the \gls{nep}-\gls{go} model are freely available at \url{https://gitlab.com/brucefan1983/nep-data}. All \gls{rgo} structures obtained from thermal reduction simulations are available in extended xyz format on Zenodo at \url{https://doi.org/10.5281/zenodo.18027655}.

\vspace{0.5cm}
\noindent{\textbf{Declaration of competing interest:}}

The authors declare that they have no competing interests.

\bibliography{refs}

\end{document}